\documentclass[twocolumn,aps,pre,superscriptaddress]{revtex4-1}

\usepackage{eurosym}
\usepackage{mathpazo}
\usepackage[T1]{fontenc}
\usepackage[latin9]{inputenc}
\usepackage{color}
\usepackage{amsmath}
\usepackage{amssymb}
\usepackage{graphicx}
\usepackage[breaklinks=true,colorlinks=true,citecolor=blue,filecolor=black,linkcolor=red,urlcolor=blue]{hyperref}

\begin{document}

\title{Double-layer Bose-Einstein condensates: A quantum phase transition in
the transverse direction, and reduction to two dimensions}

\author{Mateus C. P. dos Santos}
\affiliation{Instituto de F\'{i}sica, Universidade Federal de Goi\'{a}s 74.690-970, Goi\^{a}nia,
Goi\'{a}s, Brazil}

\author{Boris A. Malomed}
\affiliation{Department of Physical Electronics, School of Electrical Engineering,
Faculty of Engineering, Tel Aviv University, and Center for Light-Matter
Interaction, Tel Aviv University, Tel Aviv 69978, Israel}
\affiliation{Instituto de Alta Investigaci\'{o}n, Universidad de Tarapac\'{a}, Casilla 7D,
Arica, Chile}

\author{Wesley B. Cardoso}
\email{wesleybcardoso@ufg.br}
\affiliation{Instituto de F\'{i}sica, Universidade Federal de Goi\'{a}s 74.690-970, Goi\^{a}nia,
Goi\'{a}s, Brazil}

\begin{abstract}
We revisit the problem of the reduction of the three-dimensional (3D)
dynamics of Bose-Einstein condensates, under the action of strong
confinement in one direction ($z$), to a 2D mean-field equation. We address
this problem for the confining potential with a singular term, \textit{viz}%
., $V_{z}(z)=2z^{2}+\zeta ^{2}/z^{2}$, with constant $\zeta $. A quantum
phase transition is induced by the latter term, between the ground state
(GS) of the harmonic oscillator and the 3D condensate split in two parallel
non-interacting layers, which is a manifestation of the \textquotedblleft
superselection\textquotedblright\ effect. A realization of the respective
physical setting is proposed, making use of resonant coupling to an optical
field, with the resonance detuning modulated along $z$. The reduction of the
full 3D Gross-Pitaevskii equation (GPE) to the 2D nonpolynomial Schr\"{o}dinger
equation (NPSE) is based on the factorized \textit{ansatz}, with the $z$%
-dependent multiplier represented by an exact GS solution of the 1D
Schr\"{o}dinger equation with potential $V(z)$. For both repulsive and attractive
signs of the nonlinearity, the 2D NPSE produces GS and vortex states, that
are virtually indistinguishable from the respective numerical solutions
provided by full 3D GPE. In the case of the self-attraction, the threshold
for the onset of the collapse, predicted by the 2D NPSE, is also virtually
identical to its counterpart obtained from the 3D equation. In the same
case, stability and instability of vortices with topological charge $S=1$, $%
2 $, and $3$ are considered in detail. Thus, the procedure of the
spatial-dimension reduction, 3D $\rightarrow $ 2D, produces very accurate
results, and it may be used in other settings.
\end{abstract}

\maketitle

\section{Introduction}

Bose-Einstein condensates (BECs) have become a versatile platform for
realization of various phenomena, such as the production of bright \cite%
{Khaykovich_SCI02,Strecker_NAT02,Cornish_PRL06,Marchant_NC13} and dark \cite%
{Burger_PRL99} solitons, dark-bright complexes \cite{Becker_NP08}, vortices
\cite{Matthews_PRL99} and vortex-antivortex dipoles \cite%
{Neely_PRL10,Freilich_SCI10,Seman_PRA10,Middelkamp_PRA11}, persistent flows
in the toroidal geometry \cite{Ryu_PRL07,Ramanathan_PRL11,Yakimenko},
skyrmions \cite{Wuster_PRA05}, emulation of gauge fields \cite{Lin_NP11} and
spin-orbit coupling \cite{Lin_NAT11}, quantum Newton's cradles \cite%
{Kinoshita_NAT06}, Anderson localization of matter waves \cite%
{Billy_NAT08,Roati_NAT08}, rogue waves \cite{Charalampidis_RRP18}, quantum
droplets (self-trapped states supported by beyond-mean-field interactions)
\cite%
{Petrov_PRL15,Petrov_PRL16,Schmitt_NAT16,Cabrera_SCI18,Semeghini_PRL18,DErrico_PRR19,Li_PRA18,Tengstrand_PRL19,Kartashov_PRL19,Morera_PRR20}%
, etc. Further details can be found in reviews, both earlier \cite%
{Dalfovo_RMP99,Weiner_RMP99,Bloch_RMP08,BRAZHNYI_MPLB04,ABDULLAEV_IJMPB05,Gati_JPB07,Fetter_RMP09,Frantzeskakis_JPA10,Lahaye_RPP09,Song_FP13,Ueda_RPP14}
and more recent ones \cite%
{Zhai_RPP15,BAGNATO_RRP15,Lin_JPB16,Zhang_FP16,Moses_NP17,Salasnich_OQE17,Sakaguchi_FP19,Zhang_AP18,Kartashov_NRP19}%
.

Lower-dimensional BECs, i.e., bosonic gases tightly confined in one or two
transverse direction by a strong potential, make it possible to study
specific phase transitions and collective excitations in quantum settings
\cite{Bagnato_PRA91,Malomed_CM18}. In particular, studies of
quasi-two-dimensional (quasi-2D) BEC with embedded 2D potentials have drawn
much interest \cite{Bloch_RMP08,Gorlitz_PRL01}. In this connection,
approximations which make it possible to reduce the underlying 3D
Gross-Pitaevskii equation (GPE) to effective 1D \cite%
{Salasnich_PRA02,Salasnich_PRA02-2,Salasnich_PRA07-2,Maluckov_PRA08,Adhikari_NJP09,Salasnich_JPA09,Young_PRA10,Salasnich_JPB12,Salasnich_PRA13,Young-S_PRA13,Cardoso_EPJD17,Cardoso_SR17,Pendse_JPCM18,Couto_AP18,Santos_PLA19}
and 2D \cite%
{Mateo_PRA08,Salasnich_PRA09,Young_PRA10,Gligoric_PRA10,Salasnich_JPB12,Edwards_PRE12,Young-S_PRA13,Salasnich_PRA14,Cardoso_EPJD17,Kumar_PRA17}
equations have been elaborated. In particular, effective low-dimensional
equations were developed in Refs. \cite{Gerbier_EPL04}, \cite{Mateo_PRA08}, and
\cite{Salasnich_PRA09}. In the former work, the adiabatic approximation was
applied, using an appropriate analytical expression for the local chemical
potential, to eliminate the transverse dimensions and derive an effective 1D
equation that governs the axial mean-field dynamics of a strongly elongated
(cigar-shaped) BEC with repulsive interatomic interactions. On the other
hand, in Ref. \cite{Salasnich_PRA02} effective 1D and 2D time-dependent
nonpolynomial nonlinear Schr\"{o}dinger equations (NPSEs) were derived with
the help of the variational approximation, which accounts for the structure
of the condensate in the transverse directions. The use of the effective
equations with lower dimensionality is quite relevant, as such simplified
models may help one to gain deeper insight into the underlying dynamics, as
well as to reduce the cost of the computational work. In the experiment,
low-dimensional BEC can be created by means of the same technique which is
used in the 3D setting, i.e., laser cooling of atoms in magnetic and/or
optical trapping potentials \cite{Phillips_RMP98,Cabrera_SCI18}.

\begin{figure}[tb]
\centering \includegraphics[width=0.95\columnwidth]{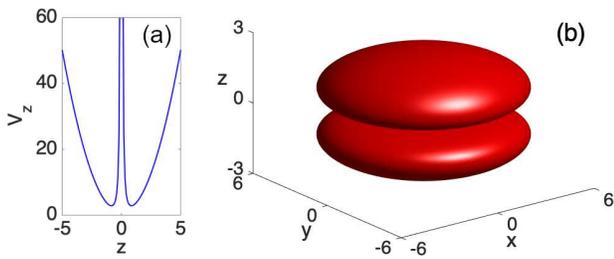}
\caption{(a) The confining transverse potential, $V_{z}=\protect\zeta%
^{2}/z^{2}+2z^{2}$, from Eq. (\protect\ref{TPOT}). (b) The 3D isosurface of
the local density, $|\protect\psi(x,y,z)|^{2}=0.003$, corresponding to the
GS (ground-state) solution of Eq. (\protect\ref{eq:GPE}) with the repulsive
sign ($g=-0.1$) of the nonlinearity and potential given by Eqs. (\protect\ref%
{TPOT}) and (\protect\ref{eq:POT1}), with $\protect\lambda=0.1$ and $\protect%
\zeta=1$.}
\label{F1}
\end{figure}

In this work, we aim to elaborate a model of BEC loaded in a 2D planar
harmonic-potential (HO) trap applied in the plane of $\left(x,y\right)$,
combined with competing HO trap $\sim z^{2}$ and singular repulsive
potential $\thicksim1/z^{2}$ acting along the perpendicular axis. This
scheme shapes the condensate into a double-pancake configuration parallel to
the plane of $z=0$, as shown in Fig \ref{F1}. Then, using the technique
similar to that elaborated in Ref. \cite{Salasnich_PRA02}, we derive an
effective 2D equation governing the system's planar dynamics. Note that,
differently from the attractive potentials $\sim-1/r^{2}$, considered in
works \cite%
{Denschlag_EPL97,Denschlag_PRL98,Sakaguchi_PRA11-2,Sakaguchi_PRA11,Astrakharchik_PRA15,Morera_PRR20}
(see also a brief review in \cite{Shamriz_CM20}), which may absorb atoms
through the mechanism of the quantum collapse (alias ``fall onto the
center'' \cite{Landau_77}), the total number of atoms is maintained
constant, in the case of the repulsive singular potential.

The rest of the paper is organized as follows. In the next section we
present the model, including a possibility of its physical realization, and
derive an effective NPSE for it. This section also contains exact analytical
results for bound states and a spectrum produced by the singular trapping
potential, and for a quantum phase transition between the HO potential and
the one which includes the singular term. In Sec. \ref{Sec3} we check the
accuracy of the effective 2D model, by comparing its predictions with
results produced by the full three-dimensional GPE. This is done for static
and dynamical states alike, including the ground state (GS) and vortex
modes. Stability and instability of vortices with topological charges $S=1$,
$2$, and $3$ is also addressed in Sec. \ref{Sec3}. The paper is concluded by
Sec. \ref{Sec4}.

\section{The effective two-dimensional NPSE (nonpolynomial Schr\"{o}dinger
equation) and analytical considerations\label{Sec2}}

\begin{figure*}[t]
\centering \includegraphics[width=0.8\textwidth]{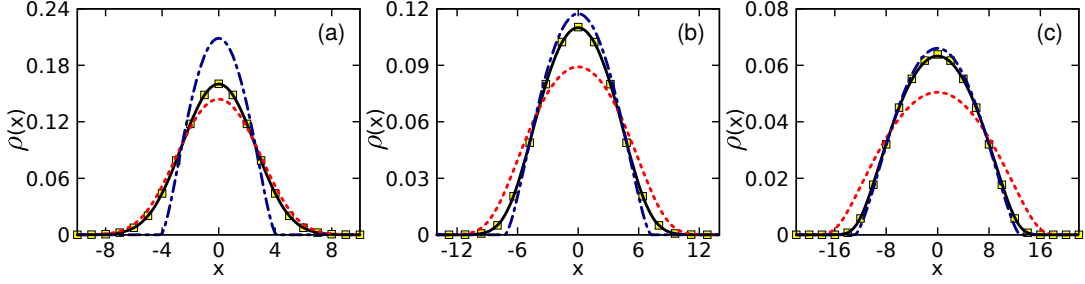}
\caption{Normalized density profile $\protect\rho (x)$ in the central cross
section (drawn through $y=0$) of the 2D GS for the repulsive BEC with
nonlinearity strength $g=1$ (a), $g=10$ (b), and $g=100$ (c), in the
presence of the in-plane potential (\protect\ref{eq:POT1}) with $\protect%
\lambda =0.1$ and $\protect\zeta =1$. The corresponding density profiles for
the full 3D GPE (\protect\ref{eq:GPE}), 2D NPSE (\protect\ref{eq:EFF}), 2D
cubic NLSE (\protect\ref{eq:W}), and TFA (\protect\ref{eq:TF}) are plotted
by chains of yellow squares, black solid lines, red dashed lines, and blue
dashed-dotted lines, respectively.}
\label{F2}
\end{figure*}

\subsection{Basic equations}

We start by considering the 3D GPE for atomic BEC, written in the usual
scaled form \cite{Pitaevskii_03,Pethick_08}:
\begin{equation}
i\frac{\partial\psi}{\partial t}=-\frac{1}{2}\mathbf{\nabla}%
^{2}\psi+V\psi+2\pi g|\psi|^{2}\psi,  \label{eq:GPE}
\end{equation}
where $\psi=\psi(x,y,z,t)$ is the mean-field wave function with the integral
norm set equal to $1$,
\begin{equation}
\iiint\left\vert \psi\left(x,y,z\right)\right\vert ^{2}dxdydz=1,  \label{1}
\end{equation}
$V(x,y,z)$ is a trapping potential, and $g=2Na_{s}/a_{z}$ is the strength of
the two-body interatomic interaction, with $N$ being the number of atoms,
while $a_{s}$ and $a_{z}$ are, respectively, the \textit{s}-wave scattering
length of atomic collisions and the confinement length of the HO potential
acting in the direction perpendicular to the system's plane, which is
adopted as the unit of length. Attractive and repulsive interactions
correspond, respectively, to $g<0$ and $g>0$.

In this work, we assume a combination of strong transverse and relatively
weak planar potentials, \textit{viz.},
\begin{equation}
V(x,y,z)=\left(\frac{\zeta^{2}}{z^{2}}+2z^{2}\right)+U(x,y).  \label{TPOT}
\end{equation}
The latter term is chosen as the isotropic HO,
\begin{equation}
U(x,y)=\dfrac{1}{2}\lambda^{2}\left(x^{2}+y^{2}\right),  \label{eq:POT1}
\end{equation}
with strength $\lambda^{2}$, while term $2z^{2}$ in the transverse potential
represents the usual one-dimensional HO trap \cite{Pitaevskii_03,Pethick_08}%
, with the strength normalized with the help of the scaling invariance of
Eq. (\ref{eq:GPE}).

The shape of the transverse potential, as defined by Eq. (\ref{TPOT}), is
displayed in Fig. \ref{F1} (a). As concerns the singular repulsive term in
the transverse part of the potential in Eq. (\ref{TPOT}), with scaled
strength $\zeta ^{2}$, it may represent a specifically designed physical
setting. Indeed, the repulsive action on cold atoms may be exerted by a
nearly-resonant blue-detuned optical field with frequency $\omega $, close
to frequency $\omega _{0}$ of atomic dipole oscillations (see, e.g., Refs.
\cite{Devlin_NJP16,Jarvis_PRL18} and references therein), the respective
interaction energy being proportional to $(\omega ^{2}-\omega _{0}^{2})^{-1}$%
. This dependence may be used to induce an effective repulsion potential,
imposing spatial modulation on $\omega _{0}^{2}$ by means of the Zeeman
effect \cite{Landau_77} in a spatially inhomogeneous dc magnetic field, or
quadratic Stark - Lo Surdo effect \cite{Landau_77} in an electrostatic
field, cf. Ref. \cite{LI_PRA13}, where an inhomogeneous field was used to
design spatially modulated dipole-dipole repulsion in BEC. In the present
context, the field should be shaped so as to make $\omega _{0}^{2}(z)=\omega
^{2}-\Omega ^{2}z^{2}$, which leads to the singular term in Eq. (\ref%
{eq:POT1}) with $\zeta ^{2}\sim 1/\Omega ^{2}$. In particular, in the case
of the Zeeman effect, the necessary spatial maximum or minimum of the
magnetic field at $z=0$ may be created by means of a solenoid shaped,
respectively, as a hyperboloid or \textquotedblleft barrel\textquotedblright
. In physical units, the value of $\zeta $ relevant for the experimental
realization is estimated as $\sim 50$ $\mathrm{\mu }$m$^{2}$. If the
realization is based on the Zeeman effect, the creation of nonuniform
magnetic field with the respective values of the variation length is
possible (see, e.g., Ref. \cite{magnetic}).

The use of the transverse potential given by Eq. (\ref{TPOT}) is relevant
for several reasons. First, as shown in detail, below, it offers a
relatively simple setting for the realization of a quantum phase transition,
and also gives rise to specific spectrum of excitations, which may be
effects interesting for the experimental realization. In addition, the
potential splits the 3D space in two uncommunicating subspaces, each one
admitting its own states, which thus suggests a possibility to create
\textquotedblleft quantum chimeras" under the action of this potential.

\subsection{The transverse wave function: analytical results and the quantum
phase transition}

The separation of the strong transverse and weak planar potentials in Eq. (%
\ref{TPOT}) suggests that the reduction 3D $\rightarrow $ 2D may be
facilitated by the adoption of the factorized \textit{ansatz} for the 3D
wave function,
\begin{equation}
\psi \left( x,y,z,t\right) =\chi (z,t)\Phi \left( x,y,t\right) .
\label{chiPhi}
\end{equation}%
First, the substitution of this \textit{ansatz} in Eq. (\ref{eq:GPE}),
keeping only the strong transverse potential and time derivative, leads to
the 1D linear Schr\"{o}dinger equation,
\begin{equation}
i\frac{\partial \chi }{\partial t}=\hat{H}_{z}\chi \equiv -\frac{1}{2}\frac{%
\partial ^{2}\chi }{\partial z^{2}}+\left( \frac{\zeta ^{2}}{z^{2}}%
+2z^{2}\right) \chi .  \label{eq:chi}
\end{equation}%
It is easy to find an \textit{exact solution} for the GS of this equation,
\begin{gather}
\chi _{0}(z,t)=\frac{2^{(1/2)(\alpha +1/2)}}{\sqrt{\Gamma (\alpha +1/2)}}%
|z|^{\alpha }\exp \left( -i\mu _{z}^{(0)}t-z^{2}\right) ,  \label{chi} \\
\alpha =1/2+\sqrt{1/4+2\zeta ^{2}},  \label{alpha} \\
\mu _{z}^{(0)}=2+\sqrt{1+8\zeta ^{2}},  \label{mu}
\end{gather}%
where $\Gamma $ is the Gamma-function, and the constant coefficient is
determined by the normalization condition,
\begin{equation}
\int_{-\infty }^{+\infty }|\chi (z)|^{2}dz=1.  \label{norm-z}
\end{equation}%
Note that, in particular, $\alpha =2$ for $\zeta ^{2}=1$, and $\alpha =3$
for $\zeta ^{2}=3$. It is worthy to note that the singular term, $\sim \zeta
^{2}/z^{2}$, in the integral which produces the GS energy,
\begin{equation}
\int_{-\infty }^{+\infty }\chi \hat{H}_{z}\chi dx=\mu _{z},  \label{H}
\end{equation}%
with Hamiltonian $\hat{H}_{z}$ defined by Eq. (\ref{eq:chi}), converges if
the wave function (\ref{chi}) is substituted in Eq. (\ref{H}).

There is another formal solution to Eq. (\ref{eq:chi}), represented by Eq. (%
\ref{chi}) with $\alpha$ replaced by
\begin{equation}
\tilde{\alpha}=1/2-\sqrt{1/4+2\zeta^{2}}  \label{alpha-}
\end{equation}
and a nominally lower eigenvalue,
\begin{equation}
\tilde{\mu}_{z}=2-\sqrt{1+8\zeta^{2}},  \label{mu-}
\end{equation}
instead of the values given by Eqs. (\ref{alpha}) and (\ref{mu}). However,
this solution is physically irrelevant, as the wave function is singular at $%
z=0$ {[}and non-normalizable at $\zeta^{2}\geq3/8$, which corresponds to
divergent integral in Eq. (\ref{norm-z}) and $\tilde{\mu}_{z}<0${]}, except
for the limit case of $\zeta=0$, when the solution based on Eqs. (\ref%
{alpha-}) and (\ref{mu-}) corresponds to the commonly known GS of the HO
potential, while the above solution, with $|z|$ replaced by $z$ in Eq. (\ref%
{chi}), produces the first excited state. Furthermore, the actual energy of
the singular state, as given by the integral expression (\ref{H}), diverges
for all $\zeta^{2}>0$, breaking the equality of the integral expression to $%
\mu_{z}$, Thus, there is a strong discontinuity (\textit{quantum phase
transition}) in the spectrum of bound states produced by Hamiltonian $H_{z}$
in Eq. (\ref{eq:chi}), as there is a jump of the GS, following the
introduction of an arbitrarily small value of $\zeta^{2}$.

The transverse potential displayed in Fig. \ref{F1}(a) seems as a structure
built of two symmetric potential wells separated by a tall barrier. This
configuration is used for the consideration of spontaneous symmetry breaking
of optical and matter waves in various nonlinear photonic and BEC settings
\cite{Malomed_13}. However, the singular potential barrier $\sim z^{-2}$ is
so strong that it splits the system in two non-communicating half-spaces (an
effect known as \textquotedblleft superselection\textquotedblright\ \cite%
{Avila-Aoki_PLA09}), therefore the consideration of Eq. (\ref{eq:GPE}) with
potential (\ref{TPOT}) on the entire axis, $-\infty <z<+\infty $, amounts to
solving the same problem on the half-axis, $0\leq z<\infty $. Indeed,
solutions (\ref{chi}) with all values $\zeta ^{2}>0$ satisfy boundary
conditions $\chi =d\chi /dz=0$ at $z=0$, therefore any two different
solutions, built independently at $z>0$ and $z<0$, may be matched at point $%
z=0$. This fact implies that, in the one hand, the symmetry-breaking
phenomenology becomes trivial in the present setting, but, on the other
hand, it opens a possibility to construct complexes of completely different
states, which would seem as \textquotedblleft quantum chimeras", cf. Refs.
\cite{chimera1,chimera2}.

In the same vein, it is interesting to compare the singular potential (\ref%
{chi}) to one featuring a more general singularity, \textit{viz}.,
\begin{equation}
V_{z}=\frac{\zeta^{2}}{|z|^{h}}+2z^{2},  \label{h}
\end{equation}
with $h>0$. It is easy to see that, at $h<2$, the expansion of the wave
function at $|z|\rightarrow0$ is
\begin{equation}
\chi\approx\mathrm{const}\cdot e^{-i\mu t}\left[1+\frac{2\zeta^{2}}{%
\left(2-h\right)\left(1-h\right)}|z|^{2-h}\right],  \label{2-h}
\end{equation}
except for the case of $h=1$, when Eq. (\ref{2-h}) is replaced by
\begin{equation}
\chi\approx\mathrm{const}\cdot e^{-i\mu t}\left[1+2\zeta^{2}|z|\ln\left(|z|%
\right)\right],  \label{h=00003D00003D00003D00003D1}
\end{equation}
The regular dependence of the expansions in Eqs. (\ref{2-h}) and (\ref%
{h=00003D00003D00003D00003D1}) on $\zeta^{2}$ means that the change $%
\zeta^{2}=0$ $\rightarrow$ $\zeta^{2}>0$ does not lead to a phase
transition. On the other hand, in the case of the singular potential (\ref{h}%
) with $h>2$ the asymptotic form of the wave function at $z\rightarrow0$ is
drastically different:
\begin{equation}
\chi\approx\mathrm{const}\cdot e^{-i\mu t}\exp\left(-\frac{2\sqrt{2\zeta^{2}}%
}{\left(h-2\right)|z|^{\left(h-2\right)/2}}\right),  \label{exp}
\end{equation}
which definitely implies that a strong phase transition takes place. Thus,
the case of $h=2$ in Eq. (\ref{h}), addressed in the present work, is a
critical one, in which the quantum phase transition commences.

Getting back to Eq. (\ref{eq:chi}), it is relevant to mention that the first
excited state of this Hamiltonian can also be found in an exact form:
\begin{gather}
\chi_{1}=\mathrm{const}\cdot|z|^{\alpha}\left(1-\frac{4z^{2}}{2+\sqrt{%
1+8\zeta^{2}}}\right)\exp\left(-i\mu_{z}^{(1)}t-z^{2}\right),  \label{chi1}
\\
\mu_{z}^{(1)}=6+\sqrt{1+8\zeta^{2}},  \label{mu1}
\end{gather}
where $\alpha$ is the same as given by Eq. (\ref{alpha}). Moreover, it is
easy to find an exact \textit{full spectrum} of eigenvalues for higher-order
excited states, with number $n$ (in the limit of $\zeta^{2}\rightarrow0$,
the spectrum carries over into energy eigenvalues of HO with odd numbers, $%
n_{\mathrm{OH}}\equiv1+2n$):
\begin{equation}
\mu_{z}^{(n)}=2\left(1+2n\right)+\sqrt{1+8\zeta^{2}},~n=0,1,2,...~.
\label{n}
\end{equation}
Note that this series of eigenvalues is \textit{equidistant}, similar to the
OH spectrum.

The formal eigenvalue given by Eq. (\ref{mu-}) also generates an infinite
series of higher-order ones, $\tilde{\mu}_{z}^{(n)}=$ $2\left(1+2n\right)-%
\sqrt{1+8\zeta^{2}}$, which are formal counterparts of OH eigenvalues
corresponding to even states, but all the respective wave functions are
singular, i.e., unphysical. Eigenvalues $\tilde{\mu}_{z}^{(n)}$ are
counterparts of those of HO with even numbers, $n_{\mathrm{OH}}\equiv2n$.

Lastly, it is also worthy to note that the 2D version of Eq. (\ref{eq:chi}),
i.e.,
\begin{eqnarray}
i\frac{\partial\chi_{\mathrm{2D}}}{\partial t} & = & -\frac{1}{2}\left(\frac{%
\partial^{2}}{\partial r^{2}}+\frac{1}{r}\frac{\partial}{\partial r}+\frac{1%
}{r^{2}}\frac{\partial^{2}}{\partial\theta^{2}}\right)\chi_{\mathrm{2D}}
\notag \\
& + & \left(\frac{\zeta^{2}}{z^{2}}+2z^{2}\right)\chi_{\mathrm{2D}},
\label{2D}
\end{eqnarray}
where $\left(r,\theta\right)$ are polar coordinates in the 2D plane,
produces exact solutions with the azimuthal quantum number, alias vorticity,
$l=0,1,2,...$ ($l=0$ corresponds to the GS):
\begin{gather}
\chi_{\mathrm{2D}}^{(l)}=\mathrm{const}\cdot r^{\eta}\exp\left(-i\mu_{%
\mathrm{2D}}^{(l)}t+il\theta-r^{2}\right),  \label{2D-S} \\
\eta=\sqrt{2\zeta^{2}+l^{2}},\mu_{\mathrm{2D}}^{(l)}=2\left(\sqrt{%
2\zeta^{2}+l^{2}}+1\right).  \label{eta}
\end{gather}
Note that the spectrum of the 2D eigenvalues, given by Eq. (\ref{eta}), is
not equidistant, unlike its 1D counterpart (\ref{n}).

\subsection{Derivation of the two-dimensional equation}

The spatial-dimension reduction 3D $\rightarrow$ 2D proceeds via the
variational approach, which is based on the Lagrangian density corresponding
to Eq. (\ref{eq:GPE}) with potential (\ref{TPOT}):
\begin{gather}
\mathcal{L}=\frac{i}{2}\left(\psi^{\ast}\frac{\partial\psi}{\partial t}-\psi%
\frac{\partial\psi^{\ast}}{\partial t}\right)-\frac{1}{2}|\nabla\psi|^{2}-%
\left(\frac{\zeta^{2}}{z^{2}}+2z^{2}\right)|\psi|^{2}  \notag \\
-U(x,y)|\psi|^{2}-\pi g|\psi|^{4}.  \label{eq:LAGR}
\end{gather}
We use the factorized \textit{ansatz} (\ref{chiPhi}), in which all the time
dependence is included in the planar wave function, $\Phi$, and the
transverse one is adopted in the form suggested by solution (\ref{chi}),
except for the phase factor, $\exp\left(-i\mu_{z}^{(0)}t\right)$:
\begin{equation}
\psi=\left[\Gamma\left(\alpha+1/2\right)\right]^{-1/2}\left(\frac{\sqrt{2}}{%
\sigma}\right)^{\alpha+1/2}|z|^{\alpha}\exp\left(-\frac{z^{2}}{\sigma^{2}}%
\right)\Phi,  \label{eq:ANS}
\end{equation}
where $\alpha$ is defined by Eq. (\ref{alpha}), and $\sigma=\sigma(x,y,t)$
is introduced as a variational parameter accounting for possible evolution
of the transverse-confinement size. Stationary states generated by \textit{%
ansatz} (\ref{eq:ANS}) imply the double-pancake shape in the 3D space, with $%
\psi(z=0)=0$, and a pair of symmetric maxima of density $\left\vert
\psi\left(x,y,z\right)\right\vert ^{2}$ at
\begin{equation}
z_{\max}^{2}=(\alpha/2)\sigma^{2},  \label{zmax}
\end{equation}
see Fig. \ref{F1} (b) as an illustration. Note also that \textit{ansatz} (%
\ref{eq:ANS}) is defined so that the transverse factor is subject to the
unitary normalization {[}cf. Eq. (\ref{norm-z}){]}, hence it follows from
Eq. (\ref{1}) that the 2D wave function also has its norm equal to $1$:
\begin{equation}
\iint\left\vert \Phi\left(x,y,t\right)\right\vert ^{2}dxdy=1.  \label{N2D}
\end{equation}

Next, by inserting \textit{ansatz} (\ref{eq:ANS}) in the Lagrangian density (%
\ref{eq:LAGR}), performing the integration in the transverse direction, and
neglecting the spatial derivatives of $\sigma$ (in the adiabatic
approximation, cf. Ref. \cite{Salasnich_PRA02}), one can derive the
corresponding effective Lagrangian:
\begin{gather}
\mathcal{L}_{\mathrm{eff}}=\frac{i}{2}\left(\Phi^{\ast}\frac{\partial\Phi}{%
\partial t}-\Phi\frac{\partial\Phi^{\ast}}{\partial t}\right)-\frac{1}{2}%
|\nabla_{\bot}\Phi|^{2}-\left(a+\frac{1}{2}\right)\times  \notag \\
\left(\sigma^{2}+\frac{1}{\sigma^{2}}\right)|\Phi|{}^{2}-U(x,y)|\Phi|^{2}-%
\frac{\pi}{2^{2\alpha}}\frac{\Gamma\left(2\alpha+1/2\right)}{%
\Gamma^{2}\left(\alpha+1/2\right)}g\frac{|\Phi|{}^{4}}{\sigma},  \label{Leff}
\end{gather}
where $\nabla_{\perp}$ is the gradient operator in Cartesian coordinates $%
\left(x,y\right)$. This expression gives rise to the Euler-Lagrange
equations:
\begin{align}
i\frac{\partial\Phi}{\partial t} & =-\frac{1}{2}\nabla_{\bot}^{2}\Phi+U(x,y)%
\Phi+\left(\alpha+\frac{1}{2}\right)\times  \notag \\
\left(\sigma^{2}+\frac{1}{\sigma^{2}}\right)\Phi & +\frac{\pi}{2^{2\alpha-1}}%
\frac{\Gamma\left(2\alpha+1/2\right)}{\Gamma^{2}\left(\alpha+1/2\right)}g%
\frac{|\Phi|{}^{2}}{\sigma}\Phi.  \label{eq:EFF}
\end{align}
\begin{equation}
\sigma^{4}-\frac{\pi}{2^{2\alpha}\left(2\alpha+1\right)}\frac{%
\Gamma\left(2\alpha+1/2\right)}{\Gamma^{2}\left(\alpha+1/2\right)}%
g|\Phi|^{2}\sigma-1=0.  \label{eq:SIG}
\end{equation}

The 2D nonpolynomial Schr\"{o}dinger equation (NPSE), Eq. (\ref{eq:EFF}), is
the main result of the derivation. It determines the density profile in the
2D plane, taking into regard effects of the transverse BEC structure.

First, it is natural to consider the low-density limit of Eq. (\ref{eq:EFF}%
), i.e., with $g|\Phi |^{2}\ll 1$ and $\sigma $ close to $1$, as it follows
from Eq. (\ref{eq:SIG}):
\begin{equation}
\sigma -1\approx \frac{\pi \Gamma \left( 2\alpha +1/2\right) }{2^{2\alpha
+1}\left( 2\alpha +1\right) \Gamma ^{2}\left( \alpha +1/2\right) }g|\Phi
|^{2}.  \label{s-1}
\end{equation}%
The small difference of $\sigma $ from $1$, given by Eq. (\ref{s-1}),
produces no contribution to NPSE (\ref{eq:EFF}) in the lowest approximation,
hence this equation amounts to the usual nonlinear Schr\"{o}dinger equation
(NLSE) with the cubic term:
\begin{align}
i\frac{\partial \Phi }{\partial t}& =-\frac{1}{2}\nabla _{\bot }^{2}\Phi
+U(x,y)\Phi +\left( 2\alpha +1\right) \Phi  \notag \\
& +\frac{\pi }{2^{2\alpha -1}}\frac{\Gamma \left( 2\alpha +1/2\right) }{%
\Gamma ^{2}\left( \alpha +1/2\right) }g|\Phi |^{2}\Phi .  \label{eq:W}
\end{align}

In the opposite high-density limit with the repulsive sign of the
nonlinearity, $g>0$, it is natural to apply the Thomas-Fermi approximation
(TFA) to the underlying three-dimensional GPE (\ref{eq:GPE}), neglecting the
kinetic-energy term (second derivatives) in it. In the absence of kinetic
energy, one has additional scale invariance, which makes it possible to fix $%
\zeta ^{2}=1$ in Eq. (\ref{TPOT}), keeping the unitary normalization of the
wave function, by defining
\begin{equation}
z=\sqrt{\zeta }z^{\prime },t=t^{\prime }/\zeta ,\psi =\zeta ^{-1/4}\psi
^{\prime },\lambda =\sqrt{\zeta }\lambda ^{\prime },g=\zeta ^{3/2}g^{\prime
}.
\end{equation}%
Then, dropping the primes, the TFA may be naturally based on the following
\textit{ansatz}:
\begin{equation}
\psi (x,y,z,t)=\frac{4}{\sqrt{3}}\left( \frac{2}{\pi }\right)
^{1/4}z^{2}\exp \left( -z^{2}-i\mu t\right) \Phi _{\mathrm{TF}},
\end{equation}%
cf. Eq. (\ref{eq:ANS}), where an expression for the 2D wave function is
derived by substituting this \textit{ansatz} in the 3D equation (\ref{eq:GPE}%
), dropping the second derivatives in it, and integrating the resultant
equation along coordinate $z$:
\begin{equation}
\left\vert \Phi _{\mathrm{TF}}\right\vert ^{2}=%
\begin{cases}
\dfrac{27\sqrt{6}}{80\sqrt{\pi }g}\left[ \mu -5-U\left( x,y\right) \right] ,
& \mathrm{at~}\mu >U(x,y)+5, \\
0, & \mathrm{at~}\mu \leq U(x,y)+5.%
\end{cases}
\label{eq:TF}
\end{equation}%
Finally, the respective chemical potential, $\mu _{\mathrm{TF}}$, can be
obtained from the normalization condition,
\begin{equation}
\iint |\Phi _{\mathrm{TF}}\left( x,y\right) |^{2}dxdy=1,  \label{NTF}
\end{equation}%
as per Eq. (\ref{N2D}).

\section{Numerical results \label{Sec3}}

\begin{figure}[tb]
\centering \includegraphics[width=0.9\columnwidth]{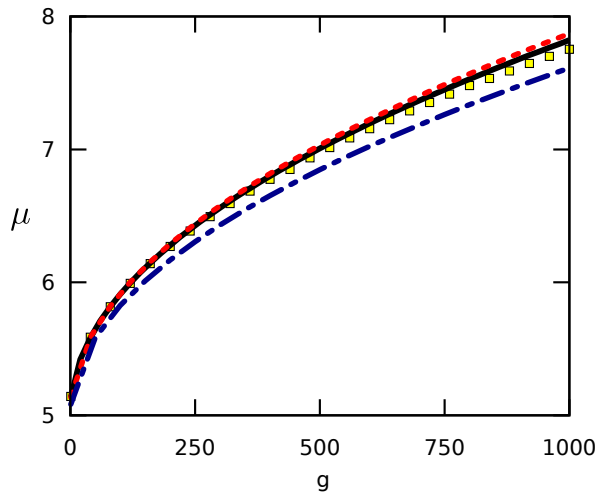}
\caption{Chemical potential $\protect\mu $ of the GS (ground state), trapped
in the in-plane potential (\protect\ref{eq:POT1}) with $\protect\lambda =0.1$
and $\protect\zeta =1$, versus the self-repulsion strength, $g$, Displayed
are numerical results obtained from the full 3D GPE, effective 2D NPSE, 2D
cubic NLSE, and TFA, under normalization conditions (\protect\ref{1}), (%
\protect\ref{N2D}) and (\protect\ref{NTF}), respectively. Symbols and curves
have the same meaning as in Fig. \protect\ref{F2}.}
\label{F3}
\end{figure}

GS solutions of the 2D and 3D equations addressed in this work were produced
by means of the well-known method of the imaginary-time evolution \cite%
{Yang_10}. It was realized by means of a split-step scheme, based on the
Crank-Nicolson algorithm, with space and time steps $\Delta_x=0.04$ and $%
\Delta_t=0.001$, respectively, in 2D and 3D cases alike. For details of
conducting numerical simulations of this type see, e.g., Refs. \cite%
{Muruganandam_CPC09,Young_CPC17}. To check the accuracy of the effective 2D
NPSE (\ref{eq:EFF}), results produced by this equation, as well as those
provided by the 2D cubic NLSE (\ref{eq:W}), and the TFA based on Eq. (\ref%
{eq:TF}), were compared to those obtained from the numerical solution of the
full 3D GPE (\ref{eq:GPE}). Below, we report the results for the GS and
vortex modes, in both cases of the repulsive and attractive nonlinearity,
i.e., $g>0$ and $g<0$.

\subsection{GS (ground-state) solutions}

\subsubsection{The repulsive nonlinearity}

\begin{figure}[tb]
\centering \includegraphics[width=1\columnwidth]{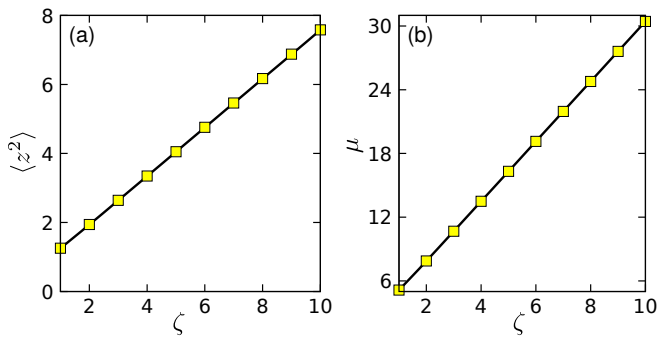}
\caption{(a) Mean-squared axial length $\langle z^{2}\rangle $ and (b)
chemical potential $\protect\mu $ of the GS versus the scaled strength $%
\protect\zeta $ of the transverse potential barrier. The  GS is trapped in
the in-plane potential (\protect\ref{eq:POT1}), with $\protect\lambda =0.1$.
The numerical results produced by the full 3D-GPE are displayed by yellow
squares, and those obtained from the effective 2D NPSE are represented by
black solid lines. The results were generated with normalization conditions (%
\protect\ref{1}) and (\protect\ref{N2D}), respectively.}
\label{F4}
\end{figure}

\begin{figure*}[t]
\centering \includegraphics[width=0.8\textwidth]{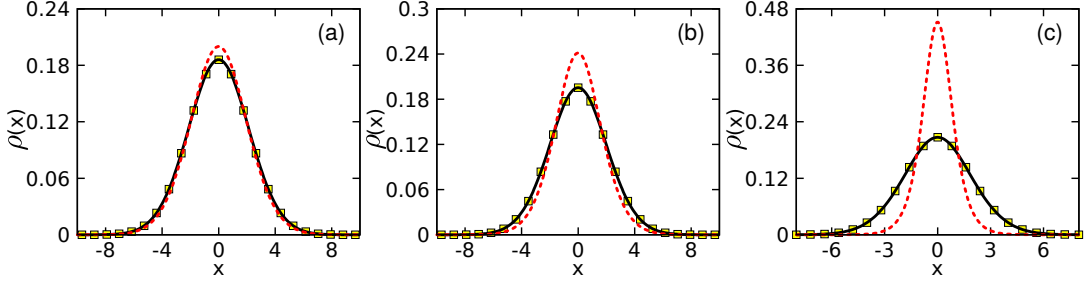}
\caption{The cross section of 2D density profiles $\protect\rho (x)$ for the
GS in the attractive condensate with $g=-0.3$ (a), $g=-0.6$ (b), and $g=-0.9$
(c), in the presence of the in-plane trapping potential (\protect\ref%
{eq:POT1}) with $\protect\lambda =0.1$ and $\protect\zeta =1$. The results
produced by the full 3D GPE, 2D NPSE, and 2D cubic NLSE are shown,
severally, by yellow squares, black solid lines, and red dashed lines.}
\label{F5}
\end{figure*}

Here, we compare the 1D lumped density profile produced by the full 3D
equation (\ref{eq:GPE}) for the stationary GS,
\begin{equation}
\rho(x)=\int\int|\psi\left(x,y,z,t\right)|^{2}dydz,  \label{eq:DENSX}
\end{equation}
with its counterpart obtained from the 2D equations (\ref{eq:EFF}), (\ref%
{eq:W}), and (\ref{eq:TF}), calculated as
\begin{equation}
\rho(x)=\int_{-\infty}^{+\infty}|\Phi\left(x,y,t\right)|^{2}dy.
\end{equation}
We start the numerical analysis by considering the repulsive nonlinearity ($%
g>0$) in the presence of the 2D (in-plane) HO trapping potential. Note that
the configuration of the BEC can be defined as double-pancake-shaped if the
transverse confinement is much tighter than the in-plane potential, i.e., $%
\lambda^{2}\ll\alpha^{-4}$, see Eq. (\ref{zmax}).

In Fig. \ref{F2} we display in-plane density profiles of the GS in the
repulsive condensate ($g>0$) under the action of potential (\ref{eq:POT1}),
for three different values of $g$. In comparison to the full 3D GPE, the 2D
NPSE provides virtually exact results, while the low-density limit and TFA,
based on Eqs. (\ref{eq:W}) and (\ref{eq:TF}), respectively, produce visible
discrepancies. In particular, for the case of relatively weak nonlinearity,
with $g=1$, displayed in Fig. \ref{F2} (a), at the central point ($x=0$) the
error between the numerically exact value of the density, obtained from the
full 3D GPE equation, and the 2D approximations is $\simeq 0.05\%$, $\simeq
10\%$, and $\simeq 30\%$ for the 2D NPSE, cubic 2D NLSE, and TFA,
respectively (naturally, TFA is irrelevant in the case of weak
nonlinearity). In the opposite case of strong nonlinearity ($g=100$), shown
in Fig. \ref{F2} (c), the same percentage errors are $\simeq 0.9\%$, $\simeq
21\%$, and $\simeq 3\%$ (in this case, the TFA is quite relevant, while the
low-density approximation is not). Note that the error produced by the 2D
NPSE increases with the increase of $g$, remaining, nevertheless, fairly
small. Similar to the situation considered in Ref. \cite{Salasnich_PRA02},
this happens because \textit{ansatz} (\ref{eq:ANS}), used for the 3D $%
\rightarrow $ 2D reduction method, is taken as a solution of Eq. (\ref{chi})
with $g=0$, thus getting less accurate with the increase of $g$.

Another way of evaluating the accuracy of the 2D NPSE is through the
calculation of chemical potential $\mu $, setting $\psi (x,y,t)=\psi
(x,y)\exp \left( -i\mu t\right) $ and $\Phi (x,y,t)=\Phi (x,y)\exp \left(
-i\mu t\right) $ in Eqs. (\ref{eq:EFF}) and (\ref{eq:W}), while, as
mentioned above, $\mu _{\mathrm{TF}}$ is defined by normalization condition (%
\ref{NTF}). Figure \ref{F3} shows that the chemical potentials obtained from
the 3D GPE and 2D NPSE always stay very close, while the low-density
approximation and TFA produce discrepancies. Note that the positive slope of
the $\mu (g)$ dependence is tantamount to $d\mu /dN>0$ , if $g$ is kept
constant, while norm $N$ is not fixed by Eq. (\ref{N2D}) but is allowed to
vary. In turn, the latter condition (\textit{the anti-Vakhitov-Kolokolov
criterion}) is necessary for stability of localized states under the action
of self-repulsion \cite{Sakaguchi_PRA10}. In fact, for simple modes, such as
GS, this criterion may be sufficient for the stability, which is confirmed
by direct simulations of their perturbed evolution (not shown here in
detail).

We also analyzed the efficiency of the 2D NPSE with respect to variation of
the transverse scaled strength $\zeta $. To this end, we set $\lambda =0.1$
and $g=1$, and calculate the mean squared axial length,
\begin{equation}
\langle z^{2}\rangle =\iiint \frac{2^{\alpha +1/2}|z|^{2(\alpha
+1)}e^{-2z^{2}/\sigma ^{2}}}{\Gamma \left( \alpha +1/2\right) \sigma
^{2\alpha +1}}\left\vert \Phi (x,y,t)\right\vert ^{2}dxdydz,
\end{equation}%
based on the \textit{ansatz} (\ref{eq:ANS}) for the 2D NPSE, which we
compare to $\langle z^{2}\rangle =\iiint z^{2}\left\vert \psi \left(
x,y,z\right) \right\vert ^{2}dxdydz$, calculated as per the 3D GPE. This
result is shown in Fig. \ref{F4}(a), where we see that the effective
equation maintains its accuracy with the increase of $\zeta $. Corroborating
this result, in Fig. \ref{F4}(b) we present the behavior of the chemical
potential $\mu (\zeta )$, in which the linear dependence $\mu (\zeta )$,
produced by the 2D NPSE, is virtually overlapped with that produced by the
3D GPE.

\subsubsection{The attractive nonlinearity}

Next, we address the case of the self-attraction, $g<0$. For this case, in
Fig. \ref{F5} we plot cross-sections of the 2D density profiles, as produced
by the full 3D GPE, 2D NPSE, and cubic 2D NLSE (the TFA is irrelevant for
the attractive nonlinearity). Similar to the case of $g>0$, the 2D NPSE
predicts the profiles in a virtually exact form, while the cubic 2D NLSE
produces a visible discrepancy with the increase of $|g|$.

Furthermore, Eq. (\ref{eq:EFF}) with $g<0$ gives rise to the collapse of the
wave function, when the strength of the self-attraction exceeds a critical
value, $|g|>|g_{c}|$, as may be expected in the 2D setting \cite{Fibich_15}.
We have found the value of $g_{c}$ for the attractive BECs in the presence
of the in-plane trapping potential (\ref{eq:POT1}), under normalization
conditions (\ref{1}) and (\ref{N2D}), respectively, with $\lambda =0.1$,
using the full 3D GPE, as well as the other approximations. We have thus
obtained $g_{c}=-2.1$ from the 2D NPSE, which is exactly the same as
produced by the full 3D GPE. On the other hand, in the framework of the
cubic 2D NLSE the collapse occurs at $g_{c}=-2.5$.

\subsection{Vortex modes under the action of the repulsive and attractive
nonlinearity}

\begin{figure*}[t]
\centering \includegraphics[width=0.8\textwidth]{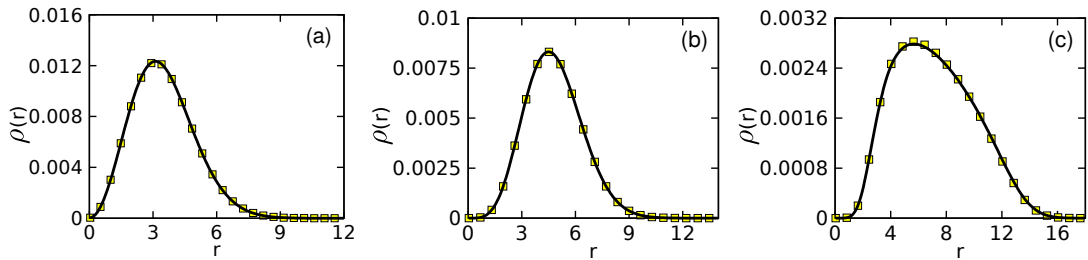}
\caption{Radial density profiles $\protect\rho (r)$ for states with
vorticity $S$ trapped in the in-plane potential (\protect\ref{eq:POT1}) with
$\protect\lambda =0.1$ and $\protect\zeta =1$, pertaining to: (a) $S=1$ and $%
g=-0.8$, (b) $S=2$ and $g=1$, and (c) $S=3$ and $g=100$. The profiles
obtained from the full 3D GPE and 2D NPSE are displayed, severally, by
yellow squares and black solid lines. }
\label{F6}
\end{figure*}

Here we address vortex solutions produced by the full 3D and effective 2D
equations. To this end, the 3D wave function is looked for, in the
cylindrical coordinates, as $\psi(r,\theta,z,t)=\Psi(r,z,t)\exp\left(iS%
\theta\right)$, resulting in the following equation:
\begin{align}
i\frac{\partial\Psi}{\partial t} & =-\frac{1}{2}\left[\dfrac{\partial^{2}\Psi%
}{\partial r^{2}}+\dfrac{\partial^{2}\Psi}{\partial z^{2}}+\dfrac{1}{r}%
\dfrac{\partial\Psi}{\partial r}\right]+\dfrac{S^{2}}{r^{2}}\Psi  \notag \\
& +\left(\frac{\zeta^{2}+2z^{4}}{z^{2}}+\dfrac{1}{2}\lambda^{2}r^{2}\right)%
\Psi+2\pi g|\Psi|^{2}\Psi,  \label{eq:GPEC}
\end{align}
where $S$ is integer vorticity. Similarly, substituting $\Phi(r,\theta,t)=%
\phi(r,t)\exp\left(iS\theta\right)$ in the 2D NPSE equation (\ref{eq:EFF})
leads to the radial equation
\begin{align}
i\frac{\partial\phi}{\partial t} & =-\frac{1}{2}\left[\dfrac{\partial^{2}\phi%
}{\partial r^{2}}+\dfrac{1}{r}\dfrac{\partial\phi}{\partial r}\right]%
+\left(\alpha+\frac{1}{2}\right)\left(\sigma^{2}+\frac{1}{\sigma^{2}}%
\right)\phi  \notag \\
& +\frac{1}{2}\lambda^{2}r^{2}\phi+\dfrac{S^{2}}{r^{2}}\phi+\frac{\pi}{%
2^{2\alpha-1}}\frac{\Gamma\left(2\alpha+1/2\right)}{\Gamma^{2}\left(%
\alpha+1/2\right)}g\frac{|\phi|^{2}}{\sigma}\phi,  \label{eq:EFFC}
\end{align}
which is combined with Eq. (\ref{eq:SIG}).

In Fig. \ref{F6}, we display examples of radial density profiles of the
vortex modes produced by Eqs. (\ref{eq:GPEC}) and (\ref{eq:EFFC}), with $%
\rho(r)=\int_{-\infty}^{+\infty}\left\vert \Psi\left(r,z,t\right)\right\vert
^{2}dz$ and $\rho(r)=|\phi(r,t)|^{2}$, respectively. The profiles are
presented for both $g<0$ and $g>0$, and for three values of the vorticity, $%
S=1,2,3$. The results clearly corroborate the accuracy of the 2D NPSE in
describing the vortex states of the 3D GPE.

\begin{figure*}[t]
\centering \includegraphics[width=0.8\textwidth]{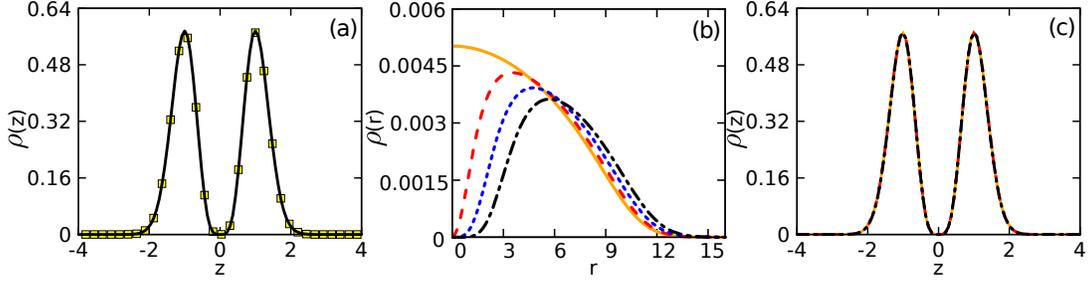}
\caption{(a) Lumped axial density profiles $\protect\rho(z)$ for the
repulsive condensate, with $\protect\lambda=0.1$, $g=1$, and vorticity $S=2$%
, produced by the full 3D GPE and 2D NPSE (yellow squares and black solid
lines, respectively. (b) Radial density profiles, $\protect\rho(r)$, for the
repulsive condensate with $g=50$, under the action of the in-plane trapping
potential (\protect\ref{eq:POT1}) with $\protect\lambda=0.1$ and $\protect%
\zeta=1$, as produced by the 2D NPSE with normalization (\protect\ref{N2D}),
for different values of the vorticity: $S=0$, $1$, $2$, and $3$. They are
depicted, respectively, by the orange solid, red dashed, blue dotted, and
black dashed-dotted lines. (c) Axial density profiles, $\protect\rho(z)$,
for the same states as in panel (b). All the profiles are practically
overlapping.}
\label{F7}
\end{figure*}

The dimensional reduction method employed here, which is based on \textit{%
ansatz} (\ref{eq:ANS}), can also be used to produce lumped density profiles
of the condensate in the axial direction, by the integration in the $\left(
x,y\right) $ plane:
\begin{equation}
\rho (z)=\frac{2^{\alpha +3/2}\pi }{\Gamma \left( \alpha +1/2\right) }%
|z|^{2\alpha }\int_{0}^{\infty }\exp \left( -2\dfrac{z^{2}}{\sigma ^{2}}%
\right) \frac{\left\vert \phi (r,t)\right\vert ^{2}}{\sigma ^{2\alpha +1}}%
rdr.  \label{eq:perf}
\end{equation}%
It is relevant to compare this approximate result to its counterpart, i.e.,
the integrated density, provided by the full 3D GPE as
\begin{equation}
\rho (z)=2\pi \int_{0}^{\infty }|\Psi \left( r,z,t\right) |^{2}rdr.
\label{eq:perfr3d}
\end{equation}%
The comparison, presented in Fig. \ref{F7}(a) for typical axial profiles
with $S=2$, again demonstrates that the 2D NPSE offers virtually exact
results, i.e., it can be reliably used for the full description of the
zero-vorticity and vortex modes. In particular, it demonstrates that the
radius of the central \textquotedblleft hole\textquotedblright\ in the
trapped mode, induced by the embedded vorticity, increases with the growth
of $S$, as can be clearly seen in Fig. \ref{F7}(b). This is a general
property of solitons with embedded vorticity, which admits an analytical
explanation \cite{Qin_PRA16}. On the other hand, the axial density profile
is very weakly affected by the value of $S$, as shown by Fig. \ref{F7}(c).

Addressing the onset of the collapse in vortex states with $S=$ $1$, $2$ and
$3$, confined by the in-plane trapping potential (\ref{eq:POT1}) with $%
\lambda =0.1$, we have computed a set of respective critical values of the
self-attraction coupling constant, $g_{c}$, for the vortices subject to
normalization conditions (\ref{1}) and (\ref{N2D}). These results were
produced by the full 3D GPE, as well as by means of the 2D NPSE and cubic
NLSE, which are collected in Table \ref{T1}, where we have also included the
result for the GS ($S=0$), presented in the previous subsection. One can see
that the 2D NPSE produces accurate predictions, in comparison to those found
from the full 3D GPE for all values of $S$. Note that $g_{c}$ strongly
increases with the growth of $S$, similar to what was observed in other
models \cite{Mihalache_PRA06}.

\begin{table}[tbp]
\begin{tabular}{|c|c|c|c|}
\hline
$S$ & $(g_{c})_{\mathrm{3D-GPE}}$ & $(g_{c})_{\mathrm{2D-NPSE}}$ & $(g_{c})_{%
\mathrm{cubic\,2D-NLSE}}$ \\ \hline\hline
$0$ & -2.1 & -2.1 & -2.5 \\ \hline
$1$ & -8.0 & -8.1 & -9.7 \\ \hline
$2$ & -12.7 & -13.2 & -18.4 \\ \hline
$3$ & -16.1 & -17.0 & -27.1 \\ \hline
\end{tabular}%
\caption{Critical values of the strength of the two-body interatomic
interaction $g_{c}$ for the onset of the collapse, calculated for the ground
($S=0$) and vortex ($S=1$, $2$ and $3$) states in the self-attractive BEC ($%
g<0$), under the action of the in-plane trapping potential (\protect\ref%
{eq:POT1}) with $\protect\lambda =0.1$ and $\protect\zeta =1$. As indicated
in the table, the critical values are calculated as per the full 3D-GPE and
2D approximations.}
\label{T1}
\end{table}

\subsection{Stability of the vortex modes with $S=1,2,3$}

To conclude the analysis of the vortex solutions under the action of the
cubic self-attraction, we studied their stability by means of direct
simulations, starting from inputs perturbed by anisotropic deformations,
which may readily initiate splitting of unstable vortices in nonlinear
models \cite{Alexander_PRE02,Saito_PRL02,Mihalache_PRA06,Malomed_PD19}. For
this purpose, we used, first, 2D NPSE (\ref{eq:EFF}) with the in-plane
trapping potential (\ref{eq:POT1}). The anisotropically deformed initial
condition with vorticity $S$ was taken as
\begin{equation}
\Phi (x,y,t=0)=\beta (x+iy)^{S}\exp \left[ -\frac{\lambda }{2}\left( x^{2}+%
\frac{y^{2}}{\gamma ^{2}}\right) \right] ,  \label{eq:INCOND}
\end{equation}%
where $\beta $ is a constant determined by the normalization condition (\ref%
{N2D}), and $\gamma $ is the parameter of the anisotropic deformation.
Below, we set $\gamma =1.1$.

\begin{figure}[tb]
\centering \includegraphics[width=1\columnwidth]{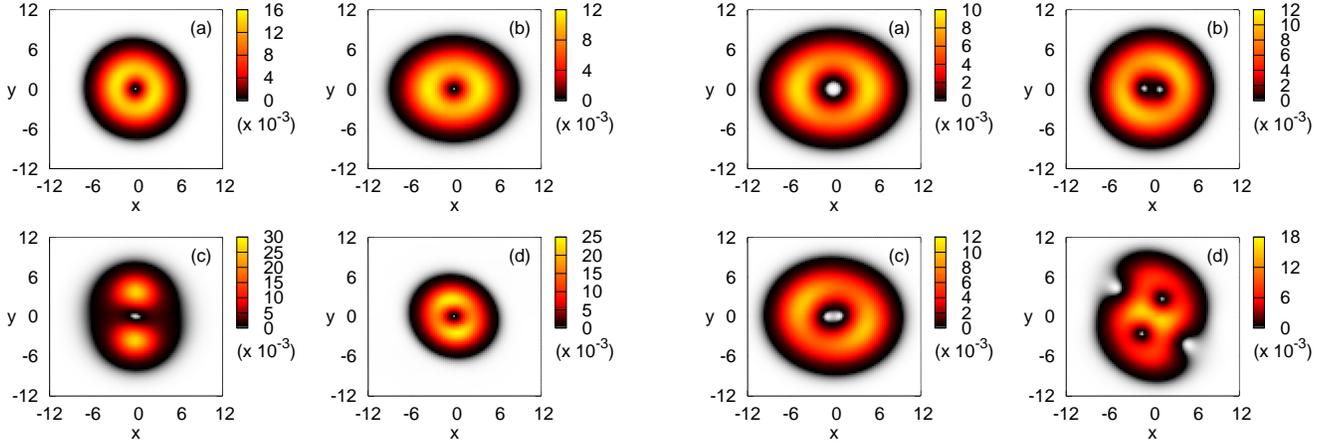}
\caption{Density profiles $|\Phi (x,y,t)|^{2}$ display the evolution of the
vortex states, with $S=1$, which were initially subjected to the elliptic
deformation, as per Eq. (\protect\ref{eq:INCOND}) with $\protect\gamma =1.1$
and $\protect\lambda =0.1$. The results were produced by simulations of Eq. (%
\protect\ref{eq:EFF}) with the in-plane potential (\protect\ref{eq:POT1})
and parameters $\protect\zeta =1$, $\protect\lambda =0.1$. (a): A snapshot,
at $t=300$, of a stable vortex profile, for $g=-1$. Panels (b), (c) and (d)
display snapshots, at $t=0$, $60$, and $140$, respectively, of the
periodically splitting and recombining vortex profile, for $g=-3.3$.}
\label{F8}
\end{figure}

From the numerical results we conclude that the vortex solutions with $S=1$
are stable at $g\geqslant -3.1$. It is worth noting that the modulus of this
value, which borders the stability region of the vortices with $S=1$, is
much smaller than the modulus of the respective value at the collapse point,
$g_{c}(S=1)=-8.1$, see Table \ref{T1}. In the stability region, perturbed
vortex states show quasi-periodic oscillations, maintaining their integrity.
These oscillations, featuring alternations of the ellipticity between the $x$
and $y$ axes (\textit{eccentricity oscillations} \cite{Christiansen_PS97}),
are caused by the initially imposed anisotropy in Eq. (\ref{eq:INCOND}). A
typical example is shown in Fig. \ref{F8}(a), where we plot a stable vortex
profile with $S=1$ and $g=-1$ obtained at $t=300$ after nine cycles of the
quasi-periodic eccentricity oscillations. These results are in agreement
with those previously published for 2D NPSE \cite{Salasnich_PRA09} and also
for cubic 2D NLSE \cite%
{Alexander_PRE02,Saito_PRL02,Mihalache_PRA06,Malomed_PD19}.

\begin{figure}[tb]
\centering \includegraphics[width=1\columnwidth]{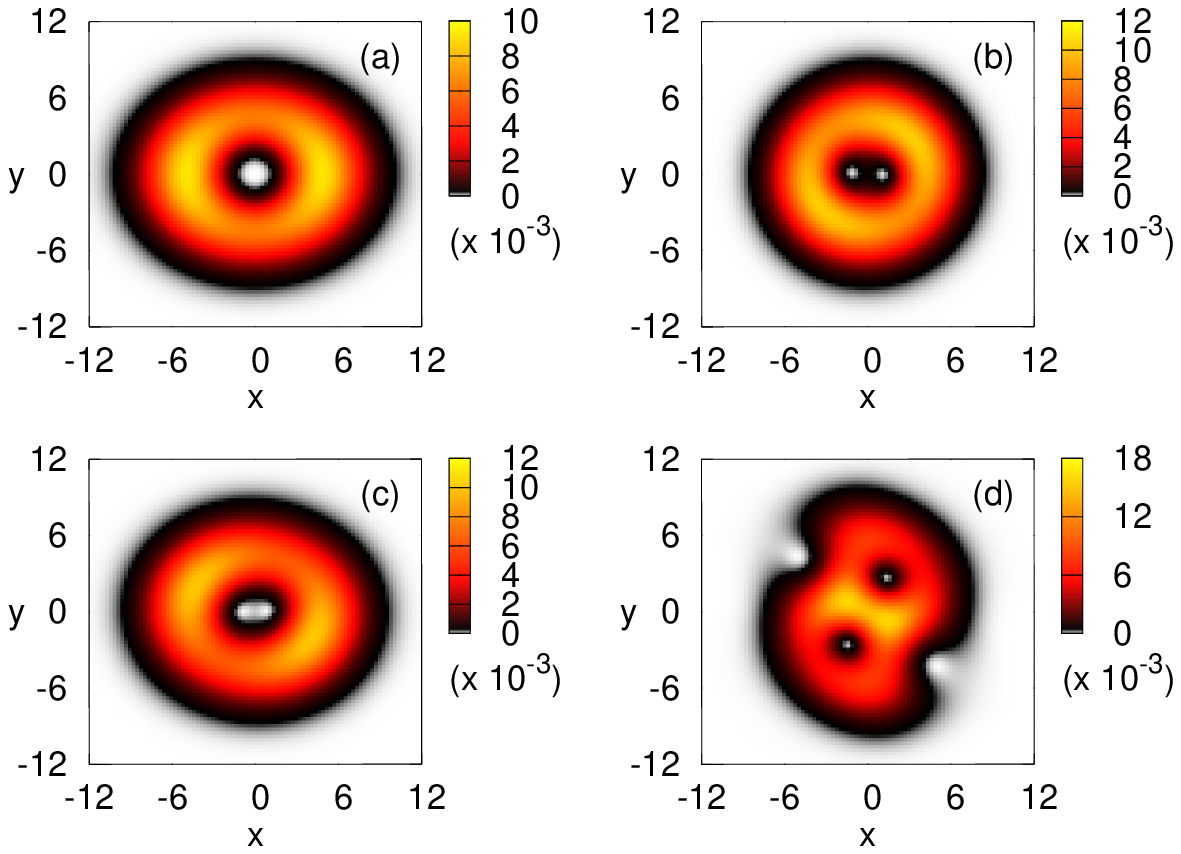}
\caption{The same as in Fig. \protect\ref{F8}, but for the vortex with $S=2$%
. Panels (a), (b) and (c) are snapshots of the solution's profile in the
quasi-periodic fission-fusion regime, taken at $t=0$, and $50$, and $100$,
respectively, for $g=-0.2$. (d) A snapshot of a permanently split (and
gradually separating) rotating two-vortex pair, taken at $t=300$ for $g=-2$.}
\label{F9}
\end{figure}

In the instability region, i.e., at $g<-3.1$, the evolution leads to fission
of the vortex ring in two fragments that rotate around the center and
recombine (fuse) again. Near the instability border, $g=-3.1$,
fission-fusion cycles repeat quasi-periodically. For example, at $g=-3.3$
the period of this dynamical regime is $\tau _{\mathrm{ff}}\simeq 140$.
These results are exhibited by means of snapshots in Figs. \ref{F8}(b-d).
For the same case, numerical data demonstrate that time necessary for the
two fragments to make a complete rotation around the origin is $\tau _{%
\mathrm{r}}\simeq 80$. Thus, the fission-fusion cycles and rotation are not
strictly synchronized, the ratio of the respective periods being
\begin{equation}
\tau _{\mathrm{ff}}\mathbf{/}\tau _{\mathrm{r}}\simeq 1.75.  \label{ff/r}
\end{equation}

In the regime of strong attractive self-interaction, the oscillation
frequency increases with the increase of $|g|$, eventually leading to the
onset of the collapse. For instance, at $g=-5$ and $S=1$, the evolution of
the initial profile (\ref{eq:INCOND}) ends up with the collapse at $%
t\simeq31 $.

For vortex states with $S=2$ no stability region was found, similar to what
was reported earlier in the case of the 2D GPE with the cubic
self-attraction and HO trapping potential \cite{Mihalache_PRA06,Malomed_PD19}%
. In the case of weak attraction, the evolution of input (\ref{eq:INCOND})
exhibits fission of the double vortex into a coupled pair of unitary
vortices with two separated pivots. Then the pair fuses into a single double
vortex, and the fission-fusion cycles for the vortices recur
quasi-periodically. Simultaneously, the configuration features persistent
rotation. This dynamical regime for the double vortex is illustrated, in
Figs. \ref{F9}(a-c), for $g=-0.2$ by dint of snapshots taken at $t=0,50,100$%
. In this case, the fission-fusion period is $\tau _{\mathrm{ff}}\simeq 31$.
It is worthy to note that the respective oscillatory motion of pivots of the
two unitary vortices proceeds along the $x$-axis, keeping $y=0$.
Simultaneously, the position of the maximum local intensity of the solution
rotates with a period of $\tau _{r}\simeq 62$. Thus, in this cases, the
ratio of the periods is
\begin{equation}
\tau _{\mathrm{ff}}\mathbf{/}\tau _{\mathrm{r}}\simeq 0.5,  \label{0.5}
\end{equation}%
quite different from the above value given by Eq. (\ref{ff/r}).

On the other hand, in the case of strong self-attraction, the double vortex
permanently splits in a pair of gradually separating unitary vortices, which
rotate around the center, unlike the dynamical scenario outlined above for
the case of weak self-attraction. In spite of the difference, for the pair
of separating unitary vortices the rotation period is found to be the same
as observed for the rotation of the maximum amplitude in case of the weak
self-attraction. The dynamics of the permanently split double vortex is
displayed in Fig. \ref{F9}(d), where local density $|\Phi (x,y,t)|^{2}$ is
plotted in the $(x,y)$ plane for $g=-2$ at $t=300$.

\begin{figure}[tb]
\centering \includegraphics[width=1\columnwidth]{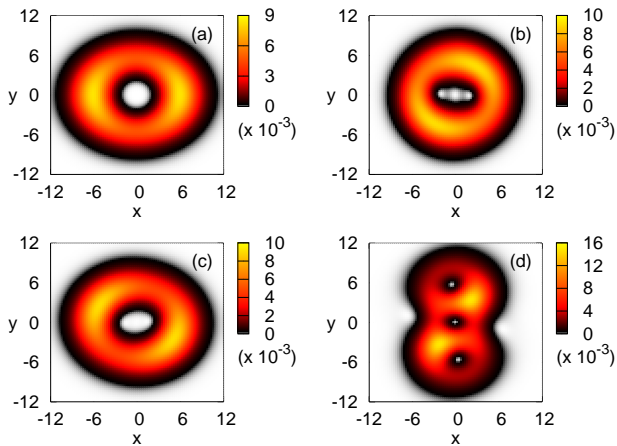}
\caption{The same as in Fig. \protect\ref{F9}, but for $S=3$.}
\label{F10}
\end{figure}

\begin{figure*}[t]
\centering \includegraphics[width=0.85\textwidth]{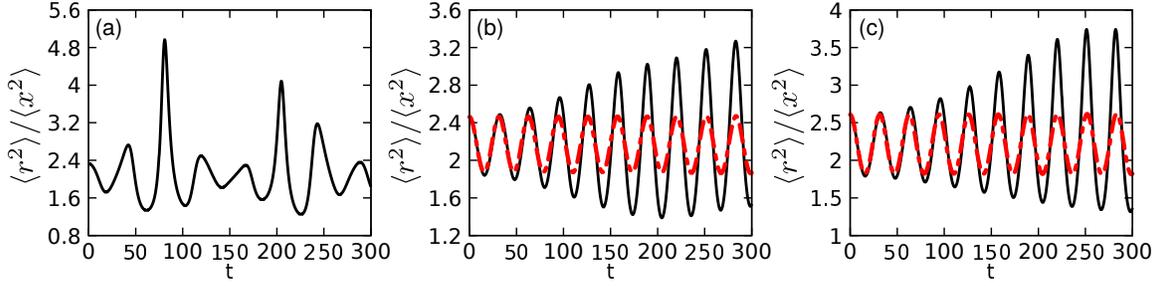}
\caption{The ratio of spatially averaged squared coordinates, $\langle
r^{2}\rangle /\langle x^{2}\rangle $ {[}see Eq. (\protect\ref{average}){]},
versus time, illustrating the unstable evolution of initial vortex states
with (a) $S=1$ and $g=-3.3$; (b) $S=2$ and $g=-0.2$ or $-2$ (the dashed red
or solid black lines, respectively; (c) the same as in (b), but for $S=3$.
The other parameters are $\protect\lambda =0.1$, $\protect\zeta =1$, and $%
\protect\gamma =1.1$.}
\label{F11}
\end{figure*}

The analysis was also developed to triple vortex states, with $S=3$,
producing results similar to those reported above for $S=2$. The triplets
are unstable against fission into a rotating set of three unitary vortices,
whose pivots are aligned in the radial direction. For small values of $|g|$,
such as $g=-0.2$, the secondary eddies temporarily fuse back into a single
triple vortex, thus initiating a quasi-periodic sequence of fission-fusion
cycles, as shown in Figs. \ref{F10}(a-c). The cycles resemble those
demonstrated above for vortices with $S=2$ in the case of the weak
self-attraction with $S=2$, with the same values of the fission-fusion and
overall-rotation periods, see Eq. (\ref{0.5}). On the other hand, permanent
splitting of the initial triple vortex in a rotating set of unitary
vortices, gradually separating in the radial direction, takes place at
larger values of $|g|$. An example of the set of three radially separating
eddies is plotted in Fig. \ref{F10}(d) for $g=-2$ at $t=300$. In such a
configuration, the separation between them values which are essentially
larger than the maximum distance observed in the fission-fusion regime. The
rotary motion of the vortex set and its period are similar to those
demonstrated in the same regime for $S=2$ in Fig. \ref{F9}(d).

Finally, to elucidate the dynamical scenarios of the instability development
outlined above for the vortex states with $S=1$, $2$, and $3$, in Fig. \ref%
{F11} we plot the evolution of ratio $\langle r^{2}\rangle /\langle
x^{2}\rangle $ of the spatially averaged 2D radial variable, $%
r^{2}=x^{2}+y^{2}$, and the squared coordinate,
\begin{equation}
\langle r^{2},x^{2}\rangle \equiv \iint (r^{2},x^{2})\left\vert \phi \left(
x,y,t\right) \right\vert ^{2}dxdy,  \label{average}
\end{equation}%
for $S=1$, $2$ and $3$ (the definition of the average does not include a
normalization factor, as it cancels in ratio $\langle r^{2}\rangle /\langle
x^{2}\rangle $). The computation was performed using numerical solutions of
the effective radial equation (\ref{eq:EFFC}) in 2D Cartesian coordinates.

In Fig. \ref{F11}(a) one observes quasi-periodic oscillations of $\langle
r^{2}\rangle /\langle x^{2}\rangle $ in an unstable state with $S=1$ and $%
g=-3.3$, as a result of the superposition of the recurring fission-fusion
cycles and overall rotation with different periods, see Eq. (\ref{ff/r}).
Note that maxima of the ratio correspond to minima of $\left\langle
x^{2}\right\rangle $, when the rotating unitary vortex is crossing axis $x=0$%
. On the other hand, in Figs. \ref{F11}(b) and (c) unstable vortices with $%
S\geq 2$ and small $\left\vert g\right\vert $ demonstrate practically
harmonic oscillations of the same ratio, $\left\langle R^{2}\right\rangle
/\left\langle x^{2}\right\rangle $, which is explained by the fact that in
this case the periods of the fission-fusion cycle and rotation are
commensurable, see Eq. (\ref{0.5}).

Finally, as shown above, unstable double and triple vortices, with $S=2$ and
$3$, at large $\left\vert g\right\vert $ irreversibly split into sets of two
or three unitary vortices displaced along the radial direction, keeping to
separate in this direction. Accordingly, the respective curves in Figs. \ref%
{F11}(b) and \ref{F11}(c) exhibit quasi-harmonic oscillations with a growing
amplitude. Eventually, the separation halts under the action of the HO
trapping potential in Eq. (\ref{eq:EFFC}).

\section{Conclusion \label{Sec4}}

The objective of this work is to produce additional results for the
important problem of the reduction of the full 3D dynamics of BEC, loaded in
an external potential, which imposes strong confinement in one direction ($z$%
), to an effective 2D form. Here, this general problem is considered for the
specific form of the confining potential, given by Eq. (\ref{TPOT}), which
is a combination of the singular repulsive term $\zeta^{2}/z^{2}$ and usual
HO (harmonic-oscillator) trap. The singular term splits the 3D condensate
into a pair of parallel non-interacting ``pancakes'', which is an example of
the ``superselection'' phenomenon. A physical realization of this
configuration is proposed, in terms of a resonant optical field, whose
frequency is subjected to appropriate modulation in direction $z$,
perpendicular to the ``pancakes''. The reduction of the underlying 3D GPE
(Gross-Pitaevskii equation) to the 2D NPSE (nonpolynomial Schr\"{o}dinger
equation) is provided by the factorized \textit{ansatz},
making use of the fact that the $z$-dependent potential admits an exact GS
(ground-state) solution of the respective Schr\"{o}dinger
equation. The full spectrum of energy eigenvalues in this transverse
potential is found in an exact form too. The potential demonstrates a
quantum phase transition between the GS (ground state) of HO in the case of $%
\zeta^{2}=0$ and the ``superselection'' state at $\zeta^{2}>0$. The
resulting two-dimensional NPSE (nonpolynomial Schr\"{o}dinger equation), with
the repulsive or attractive nonlinearity, produces GS
and vortex-state solutions, and the threshold for the onset of the collapse,
which are virtually identical to their counterparts obtained from the
numerical solution of the underlying 3D GPE. On the other hand, the 2D NLSE
with the usual cubic nonlinearity, as well as the TFA (Thomas-Fermi
approximation), give rise to conspicuous discrepancies, in comparison to the
full 3D solution. Thus, the results demonstrate high accuracy of the
appropriately formulated spatial-dimensionality reduction. This method may
be applied to other settings as well.

In particular, the existence of stable vorticity states with $S=1$ in the
case of the self-attractive sign of the nonlinearity is demonstrated by
direct simulations of the effective time-dependent 2D NPSE. On the other
hand, all higher-order vortex states with $S=2$ and $3$ are unstable. At
relatively weak self-attraction strength, the instability triggers a
quasi-periodic sequence of fission-fusion cycles, while stronger
self-attraction irreversibly splits double and triple vortices in linearly
arranged rotating sets of gradually separating unitary vortices.

As an extension of the present work, it may be interesting to consider more
sophisticated patterns, such as necklace-shaped ones built of fundamental or
vortex solitons, cf. necklace states found in various other models \cite%
{Soljacic_PRL98,Desyatnikov_PRL01,Kartashov_PRL02,Grow_PRL07}. Another
relevant extension is to consider dark solitons in the model with the
self-repulsive nonlinearity, and without the axial confinement. Such modes 
can be created by means of the well-known phase-imprinting technique \cite{imprint}.
In particular,
it may be interesting to consider moving dark solitons and collisions
between them.

\section*{Acknowledgments}

The authors acknowledge financial support from the Brazilian agencies CNPq
(\#425718/2018-2 \& \#306065/2019-3), CAPES, and FAPEG (PRONEM
\#201710267000540 and PRONEX \#201710267000503). This work was performed
under auspices of the Brazilian National Institute of Science and Technology
(INCT) for Quantum Information (\#465469/2014-0). The work B.A.M. is
supported, in part, by the Israel Science Foundation through Grant No.
1286/17, and by CAPES (Brazil) through program PRINT, grant No.
88887.364746/2019-00.

\end{document}